\documentclass{article}

\usepackage{microtype}
\usepackage{graphicx}
\usepackage{caption}      % For customizing captions
\usepackage{subcaption}   % For subfigures
\usepackage{booktabs} % for professional tables
\usepackage{multirow}

\usepackage{bm}

\usepackage{hyperref}

\usepackage[accepted]{icml2025}

\usepackage{amsmath}
\usepackage{amssymb}
\usepackage{mathtools}
\usepackage{amsthm}

\usepackage[capitalize,noabbrev]{cleveref}

\theoremstyle{plain}

\theoremstyle{definition}

\theoremstyle{remark}

\usepackage{acronym}

\acrodef{MCL}{Multiple Choice Learning}
\acrodef{PLP}{Predominant Local Pulse}
\acrodef{MCL}{Multiple Choice Learning}
\acrodef{MIR}{Music Information Retrieval}
\acrodef{SSL}{Self-Supervised Learning}
\acrodef{DBN}{Dynamic Bayesian Network}
\acrodef{OSF}{Onset Strength Function}
\acrodef{WTA}{Winner-Takes-All}
\acrodef{BCE}{Binary Cross Entropy}
\acrodef{KD-MHL}{Knowledge-Driven Multiple Hypothesis Learning}

\usepackage[textsize=tiny]{todonotes}

\icmltitlerunning{Self-Supervised Learning with Knowledge-Driven Multiple Hypothesis}
\begin{document}

\twocolumn[
% \icmltitle{Self-Supervised Learning with Knowledge-Driven Multiple Hypothesis Learning}
% \icmltitle{Domain-Informed Multi-Hypothesis Learning Provides Effective \\
% Self-Supervised Rhythm Analysis}
\icmltitle{Controlling Contrastive Self-Supervised Learning with Knowledge-Driven Multiple Hypothesis: Application to Beat Tracking}



\begin{icmlauthorlist}
\icmlauthor{Antonin Gagneré}{ltci}
\icmlauthor{Slim Essid}{ltci}
\icmlauthor{Geoffroy Peeters}{ltci}

\end{icmlauthorlist}

\icmlaffiliation{ltci}{LTCI Télécom paris}

\icmlcorrespondingauthor{Antonin Gagneré}{antonin.gagnere@telecom-paris.fr}


\vskip 0.3in
]



\printAffiliationsAndNotice{}  % leave blank if no need to mention equal contribution

\sloppy

\begin{abstract}

Ambiguities in data and problem constraints can lead to diverse, equally plausible outcomes for a machine learning task.
In beat and downbeat tracking, for instance, different listeners may adopt various rhythmic interpretations, none of which would necessarily be incorrect.
To address this, we propose a contrastive self-supervised pre-training approach that leverages  multiple hypotheses about possible positive samples in the data.
Our model is trained to learn representations compatible with different such hypotheses, which are selected with a knowledge-based scoring function to retain the most plausible ones.
When fine-tuned on labeled data, our model outperforms existing methods on standard benchmarks, showcasing the advantages of integrating domain knowledge with multi-hypothesis selection in music representation learning in particular.

\end{abstract}
\section{Introduction}
\label{intro}

Representation learning has proven effective for extracting meaningful features from complex data, with a recent focus on \ac{SSL}.
\ac{SSL} methods rely on pretext tasks—such as predicting masked inputs \cite{Devlin2019BERTPO, He_2022_CVPR} or contrasting augmented views of the same sample \cite{oord2019representation, simclr}—to learn general-purpose representations without labeled data.
This flexibility reduces the need for costly annotations and expands the applicability of learning-based approaches to domains with limited labeling resources. %\todo{rewrite this}

\begin{figure}[ht!]
  \centering
  \makebox[\linewidth][c]{%
    \includegraphics[width=1.\linewidth]{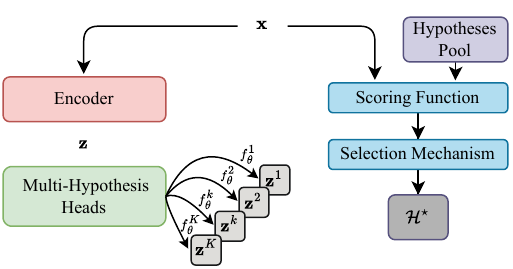} % Replace with your PDF filename
  }
  \caption{Overview of our SSL framework based on KD-MHL.
  The input sample $\bf{x}$ is encoded by $g_{\theta}$ into a representation $\bf{z}$ which is further projected by $K$ heads $f_{\theta}^k$ corresponding to multiple hypotheses $\mathcal{H}_k$ from a pool $\mathcal{H}$. Hypothesis are driven by the knowledge of the domain and lead to specific strategies for sampling anchors, positives, and negative samples within a contrastive framework. A function $h_k$ scores each hypothesis. These scores are used by a mechanism $s$ which selects the $n$ winning hypotheses. At each step, the encoder is trained considering only the winning hypotheses (i.e. considering only the loss contributions from the winning heads).}
  \label{fig:overview}
\end{figure}

This work proposes to control \ac{SSL} contrastive pre-training using multiple knowledge-driven hypotheses in the process of mining the positive and negative samples that will be considered in the contrastive scheme.
We name this framework \ac{KD-MHL}.

\ac{KD-MHL} incorporates ideas from \ac{MCL}, a learning paradigm tailored for ambiguous tasks,
where multiple outputs are estimated by an ensemble of prediction heads plugged at the output of a deep network.
Building on available annotations, various update mechanisms have been proposed to train the set of winning predictors~\cite{mcl_guzman, Rupprecht2016LearningIA, NIPS2016_20d135f0, Makansi_2019_CVPR, rmcl}.
Resilient \ac{MCL} \cite{rmcl}, in particular, jointly trains both the winning hypothesis and the scoring functions associated with each predictor. %\todo{polish this}
Without having access to annotations one cannot rely on trainable scoring heads.
Therefore \ac{KD-MHL} uses a scoring function, based on prior-knowledge, which estimates how likely it is for a training sample to correspond to a given hypothesis.
We instantiate our general framework to learn representations for musical rhythm analysis tasks,  and show that they are highly effective for the problem of automatic beat and downbeat tracking, a key task in \ac{MIR}.
It aims to identify the temporal locations of beats and downbeats in musical excerpts. The task is particularly challenging, both owing to the immense diversity of musical genres and the  ambiguity inherent in the definition of ground-truth beat/downbeat labels, where different hypotheses may sometimes be made about correct beat positions by different human annotators.

Additionally, we explore self-training as a limit case of our framework, where the selection mechanism is nearly always correct. 
Models pre-trained in this way achieve state-of-the-art performance on most benchmark datasets, often surpassing previous systems by 2\% in reference evaluation metrics. 

Our key contributions are the following: 
\begin{enumerate}
\vspace{-1em}
\setlength\itemsep{-0.3em}
\item We introduce a new \ac{SSL} framework that leverages  multiple hypothesis learning to define its contrastive scheme, accounting for task ambiguity and leading to more powerful representations which readily accommodate multiple possible outcomes during downstream task adaptation.
\item We instantiate this framework to learn representations for musical rhythm analysis tasks and successfully apply them to beat and downbeat tracking.
\item We introduce a self-training variant of our approach, which achieves state-of-the-art beat and downbeat tracking performance on almost all reference benchmark datasets.
% \vspace{-0.5em}
\end{enumerate}

\section{Related Work}
\label{related_work}

\paragraph{Multiple Choice Learning} Introduced in \cite{mcl_guzman}, \ac{MCL} is a framework that employs multiple prediction heads to address tasks with inherent ambiguity. \cite{NIPS2016_20d135f0} extended this approach to deep learning, where heads are typically trained a \ac{WTA} strategy, updating only the head that produces the best prediction. To mitigate overconfidence, \cite{Rupprecht2016LearningIA} proposed a variant that also updates non-winning heads with scaled gradients. \cite{Makansi_2019_CVPR} propose an evolving top-$n$ \ac{WTA} strategy, updating the top $n$ predictions instead of just one. To improve uncertainty modeling, \cite{rmcl} incorporated a scoring function trained alongside the heads to estimate the probability of a hypothesis being among the winners.

\paragraph{Self-Supervised Learning} 

Most common techniques for self-supervised learning leverage Siamese networks and rely on pairs of positive samples that share semantic information~\cite{Siamese}.
These pairs are artificially created from unlabeled inputs by masking them~\cite{Assran2022,BEIT,IJEPA} or by applying semantic-preserving data augmentations to them~\cite{simclr,byol,VICReg,BarlowTwins}.
The Siamese architecture is then trained to map those pairs to close locations in a latent space.

However, without additional constraints, the network would discard all information from the input and return always the same output. This phenomenon is called \emph{representation collapse}.
To prevent this, the most common strategy is to sample negative samples and repel them via a contrastive loss~\cite{simclr}. Another approach is to explicitly penalize collapsed solutions by incorporating additional loss terms, as proposed in \cite{Wang2020a,BarlowTwins,VICReg}.
%Interestingly, \cite{Garrido2023} show that there is a duality between those two approaches.

Finally, another strategy consists of breaking the asymmetry between the two branches of the Siamese network, usually leveraging a momentum encoder~\cite{byol,SimSiam,DINO}.
These techniques, while not explicitly avoiding representation collapse, have been shown to prevent this phenomenon in practice.\\

Self-supervised learning and in particular contrastive learning have been widely used in the audio domain.
\cite{oord2019representation} applies it to raw waveforms for autoregressive prediction in the latent space, and \cite{w2v} extends it to audio masked modeling, taking inspiration from \cite{Devlin2019BERTPO}.
Later, inspired by the success of SSL with Siamese networks in the computer vision domain, \cite{Saeed2020ContrastiveLO} adapts SimCLR \cite{simclr} to the audio domain by using chunks from the same audio clip as positive pairs.
Similarly, most SSL approaches originally introduced for computer vision have been then successfully applied to the audio domain, notable examples being \cite{AudioBarlowTwins,BYOLA}.

\paragraph{Self-Supervised Learning for musical applications}
In the music domain, contrastive learning has been successfully applied to music representation learning. \cite{SpijkervetB21} leverage musically relevant data augmentations to create positive pairs, and \cite{mule} scale up this approach to large music databases, achieving state-of-the-art performances in various downstream tasks.
Many application of \ac{SSL} to \ac{MIR} relies on signal processing augmentations, that alter data with a semantic meaning \textit{e.g} pitch shifting, time-stretching.
\cite{spice} trained a network to predict the known difference between two pitch-shifted version of the same audio. 
This paved the way to other equivariant \ac{SSL} \cite{dangovski2022equivariant} applications to various task such as pitch estimation \cite{riou2023pesto}, tempo detection \cite{quinton,10448098,antonintempo} or tonality estimation\cite{kong_stone,kong2025skeyselfsupervisedlearningmajor}.
Some works have explored self-supervised learning (SSL) for beat tracking.
Zero-Note Samba \cite{desblancs2023zero} pre-trains a model by training an encoder to synchronize the latent representations of percussive and non-percussive stems.
\cite{gagnere:hal-04768296} is the closest to our work. Hypothesizing a binary meter, they contrast latent representations separated by a power of two of \ac{PLP} peaks.
% \todo{clearer here}  

\begin{figure*}[ht!]
  \centering
  \makebox[\textwidth][c]{%
    \includegraphics[width=1.\textwidth]{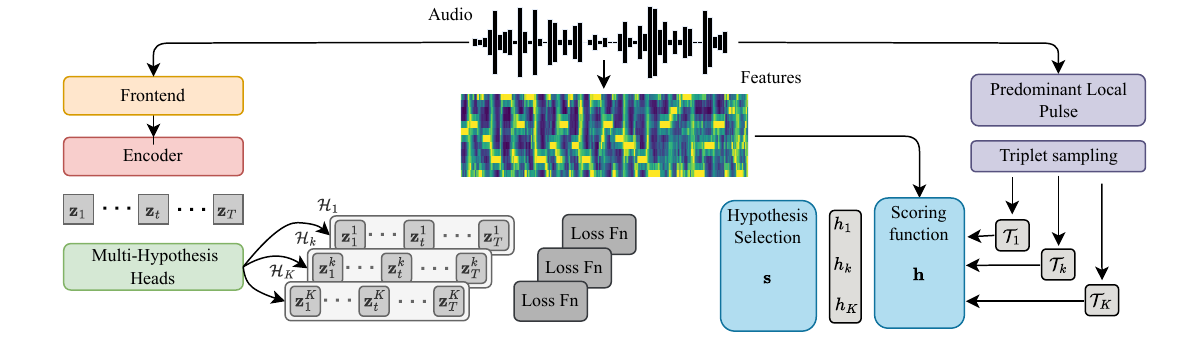} % Replace with your PDF filename
  }
  \caption{Instantiation of our SSL framework based on \ac{KD-MHL} for musical rhythm analysis (beat and downbeat tracking). 
  The input is a sequence $\bf{x}_t$ that represents the audio signal over time $t$ which is projected by $g_{\theta}$ into a sequence of $\bf{z}_t$. 
  The objective is to train $g_{\theta}$ such that $\bf{z}$ takes different values when $t$ is a beat or not.
  This is achieved using contrast learning, sampling triplets (anchors, positive and negative times). 
  Driven by knowledge, we create a pool of hypothesis $\mathcal{H}_k \in \mathcal{H}$ which correspond to possible metrical relationship between \ac{PLP} peaks  (which define the time units $t$) and beats; and therefore correspond to specific triplet samplings $\mathcal{T}_k$.
  Each $\mathcal{H}_k$ is scored by $h_k$ considering the audio features evolution under the given metrical relationship.
  These scores are used by a mechanism $s$ which selects the $n$ winning hypotheses. 
  At each step, the encoder $g_{\theta}$ is trained considering only the $n$ winning hypotheses (i.e. considering only the loss contributions from the $n$ winning heads)}.
  \label{fig:applied}
\end{figure*}
\section{Method}

\paragraph{Beat-Tracking} Data-driven approaches to beat tracking began with \cite{bocklstm}, which employed bi-directional Long Short-Term Memory (LSTM) networks to process spectral features.
This was subsequently improved by replacing LSTM with Temporal Convolutional Networks \cite{bocktcn1} and solving jointly beat and downbeat, as well as tempo detection \cite{Bck2019MultiTaskLO,Bck2020DeconstructAR}.
\cite{Hung2022} applied a Spectral-Temporal Transformer \cite{Lu2021SpecTNTAT} processing harmonic features and a temporal aggregation separately.
Beat Transformer \cite{ZhaoXW22} incorporates dilated self-attention to capture long-range dependencies alongside instrument-wise self-attention conducted along the stems of demixed audio.
All these methods rely on \ac{DBN} \cite{Krebs2013RhythmicPM} to post-process activations.
\cite{foscarin2024beatthis} achieved state-of-the-art performance by introducing shift-tolerant \ac{BCE} to remove the \ac{DBN} dependency.

An overview of our framework is presented in Figure \ref{fig:overview}.  In the following, we detail its general formulation and key elements, especially how it models in a self-supervised manner different sampling strategies using an ensemble of output heads.

\subsection{General Framework}
\label{sec:gen}

Given an input sample $\mathbf{x}$---which may be a sequence or a bag of samples from the training dataset $\mathcal{X}$---, we extract an intermediate representation $\mathbf{z}$ with an encoder $g_\theta$.
Whereas standard contrastive learning approaches typically rely on data augmentation or sample mining to define positive and negative pairs, our method specializes $\mathbf{z}$ into $K$ different representations $\{\mathbf{z}^k\}$, using an ensemble of heads $f_\theta \triangleq \bigl(f^1_\theta,\dots,f^K_\theta\bigr)$. 
Each head $f^k_\theta$ corresponds to a domain-motivated \emph{hypothesis} (driven by knowledge), $\mathcal{H}_k$, which specifies a strategy for sampling positives and negatives within the sequence or bag of samples.
We denote the set of all hypotheses by $\mathcal{H}$.

This formulation is motivated by the inherent ambiguity that may arise when predicting an output for a given task. 
Rather than forcing the model to commit to a single valid interpretation we exploit multiple heads that enable it to account for different possible outcomes.
To exploit such a method, one needs to have prior domain knowledge to define a set of plausible hypotheses but also to design the associated mining strategies. 

To guide which hypotheses should apply to $\mathbf{x}$, we introduce a scoring function $\mathbf{h}\triangleq(h_1,\ldots,h_k): \mathcal{X} \to \mathbb{R}^K$ that measures the ``compatibility'' between the data sample and each hypothesis $k$.
% This scoring function is typically knowledge-driven. \todo{better formulation here}
Let $\mathcal{T}_k = \{A_k,\;P_k,\;N_k\}$
denote, respectively, the sets of anchors, positives, and negatives under hypothesis $k$.
A selection mechanism $\bf{s} : \mathbb{R}^K \to 2^{\{1,\dots,K\}}$ outputs the subset of winning hypotheses.
% Formally, let denote by $A_k$, $P_k$, and $N_k$ denote the sets of anchors, positives, and negatives under hypothesis $k$.
The overall contrastive loss $\mathcal{L}(\mathbf{x})$ (used for \ac{SSL} training) is then defined as the sum of the contributions under the selected hypotheses:
\begin{equation}
    \label{eq:multi_hypothesis_loss}
    \mathcal{L}(\mathbf{x}) \;=\; 
    \sum_{k \,\in\, \bf{s}\bigl(\bf{h}(\mathbf{x})\bigr)}
    \mathcal{L}\Bigl(\mathbf{z}^k, \mathcal{T}_k\Bigr),
\end{equation}

By allowing multiple hypotheses to drive the choice of anchor, positive, and negative sets, we believe we help the model handle potentially ambiguous tasks.

\subsection{Hypothesis selection}

Given the score $h_k$ computed for each hypothesis $\mathcal{H}_k \in \mathcal{H}$, the selection mechanism, $\bf{s}$, determines a suitable subset of hypotheses to retain: $\mathcal{H}^{\ast}$.
We explored several selection strategies, including the \ac{WTA} approach, where only the best-scoring hypothesis is selected
% $s_{\text{WTA}} = \underset{k}{\arg\min} \ h_k(\bm{x})$,
and $n$-WTA where the $n$ best hypotheses are retained:

\begin{equation}
\mathbf{s}_{n\text{-WTA}} = \underset{\substack{\mathcal{S} \subset \{1, \dots, K\} \\ |\mathcal{S}| = n}}{\arg\min} \sum_{k \in \mathcal{S}} h_k
\end{equation}

\section{Instantiation on Musical Rhythm Analysis}

In the following, we specify the set of hypotheses $\mathcal{H}$ and their selection $\bf{h}$/$\bf{s}$ for the case of rhythm analysis (beat and downbeat tracking).
Figure \ref{fig:applied} depicts the system adapted rhythm analysis.

\subsection{Hypothesis definition}

\begin{figure}[h]
    \centering
    
    % Subfigure 1
    \begin{subfigure}[b]{\linewidth}
        \centering
        \includegraphics[height=1.8cm, width=\linewidth]{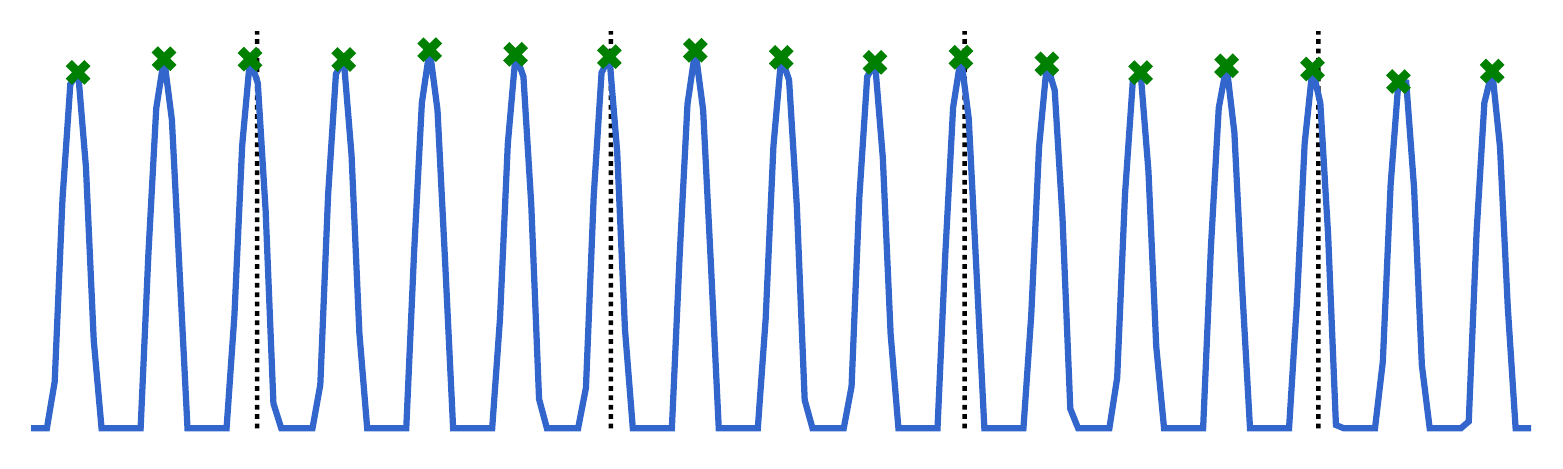} % Replace with your image path
        \label{fig:subfig1}
        % \caption{(a)}
    \end{subfigure}
    
    \vspace{-0.2cm} % Reduced vertical space between subfigures
    
    % Subfigure 2
    \begin{subfigure}[b]{\linewidth}
        \centering
        \includegraphics[height=1.8cm, width=\linewidth]{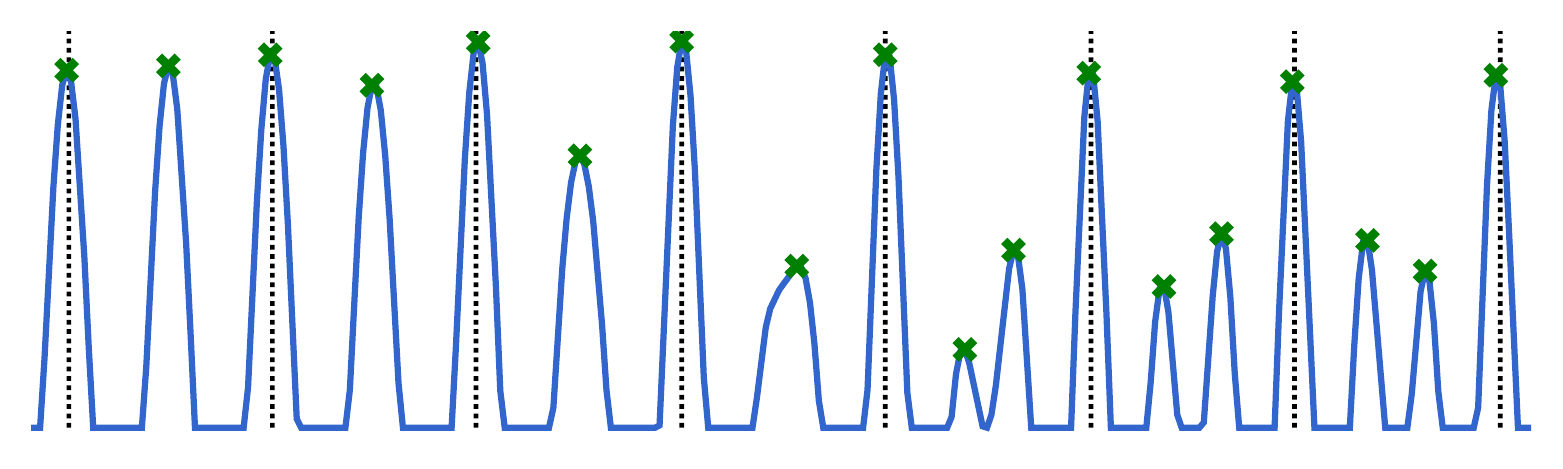} % Replace with your image path
        \label{fig:subfig2}
        % \caption{4\_2 grid example.} % Add a caption for (a)
    \end{subfigure}
    % \vspace{-0.2cm} % Reduced vertical space between subfigures
    % Main caption
    \caption{\ac{PLP} function (blue) alongside beat annotations (dashed lines) and detected peaks (green crosses). The top plot display a (4,2) hypothesis, the bottom one displays phase shifting issue.}
    \label{fig:plp}
\end{figure}

In the case of \ac{SSL}, one does not have access to ground-truth output labels for training (here the beat/downbeat positions, for instance). 
Instead, one needs to solve a pretext-task, obtaining supervision from the input data itself. 
For rhythm analysis (especially beat and down-beat detection), we rely on audio signal processing techniques to get such supervision: for each audio sample  $\mathbf{x}$ we compute the so-called predominant local pulse (PLP) function to drive the selection of plausible output hypotheses. 
This scalar function captures the dominant rhythmic pulse at each time instant in an audio signal. It is computed by first deriving a sinusoidal kernel for each time position that best explains the local periodicity of an \ac{OSF}.
The \ac{OSF} measures the likelihood that a musically salient change (\textit{e.g.}, note onset)  has occurred at each time point. All kernels are then accumulated over time using overlap-add synthesis. 
The result is a function like the one given in Figure~\ref{fig:plp}, where each peak is a plausible candidate for a beat position.

The sequence of peaks from the \ac{PLP} function is represented as a set of time positions: $\mathcal{B} = \{ b_i \}_{i=1}^{B}$, where $B$ corresponds to the number of detected peaks.
We make the assumption that these \ac{PLP} peaks $\mathcal{B}$ are aligned with the tatums.\footnote{The tatum corresponds to the fastest regular rhythmic events that occur in a piece of music. A beat interval is an integer number of tatum intervals.}
%\todo{define what is tatum; perhaps a figure would help indicating what is a PLP function, the peaks, the tatum, the beat positions}
Because of this, the actual beats form a subset of $\mathcal{B}$.
However, the exact metrical relationship between these detected peaks and the true beats is unknown.
In other words, one does not know if every peak corresponds directly to a true beat, or if the true beats only occur every $\omega$ peaks, $\omega \in \mathbb{N}$.
Also one does not know where the first true beat starts in the sequence of $\mathcal{B}$.
To formalize this, we define a ratio $\omega \in \mathbb{N}$ and an initial phase $\phi \in \mathbb{N}$ such that the subset:
%\todo{reduce size between equation}
% \add[GP]{$\omega,\phi \in \mathbb{N}$ ?} 
% \vspace{-1mm}
\begin{equation}
    \label{eq:subseq}
  \mathcal{B}_{\omega, \phi}  = \left\{ b_{\omega k + \phi} \mid 1 \leq \omega k + \phi \leq B  \right\} \; ;  \; k \in \mathbb{N}
\end{equation}
% \add[GP]{what is $M$ here ?}
corresponds to the true underlying beat sequence.
In the top part of Figure~\ref{fig:plp} the underlying ground-truth corresponds to the subset $\mathcal{B}_{4,2}$.
We denote by $\Omega$ the set of ratios considered  for $\omega$. 
Given the problem definition, the set of possible phases for a given $\omega$ is $\Phi_{\omega} = \{ \phi \mid 0 \leq \phi < \omega \}.$
This yields a total of  $K$ hypotheses with $K = \sum_{\omega \in \Omega} \omega$ :
\begin{equation}
  \mathcal{H} = \left\{\mathcal{H}_{\omega,\phi} \mid \omega \in \Omega, \phi \in \Phi_{\omega}\right\}.
\end{equation}

% We restrained the set of ratios to $\Omega = \{1, 2, 3, 4\}$.
% For each $\omega \in \Omega$, we consider the possible phases $\Phi_{\omega} = \{ k \mid 0 \leq k < \omega \}$. 
% This yields a total of $K = \sum_{\omega \in \Omega} \omega = 10$ subset. We denote these hypotheses collectively as

% To manage the hypotheses selection, i.e the one that we will train we define a knowledge informed scoring function $h$. This function is a proxy for the scoring head traditionally use in \ac{MCL} methods. 
% With the score from $h$ we can use different strategies to select the sub level representation that we will backpropagate with:

% Given an audio signal $ \mathbf{x} = \{x_l\}_{l=1}^{L}$ sampled at 16 kHz, we compute the scalar \acf{PLP} function\cite{plpGrosche}, $\{y_t\}_{t=1}^{T}$, with a hop size set to 320 samples. Using the \textit{find\_peaks} function from scipy 

\subsection{Scoring function}
\label{sec:sampling}
% Given set of hypotheses we aim to distinguish the underlying one. 
% For this we compute a contrastive scoring function on features extracted from the musical segment whose goal is to measure the mutual information between positive element compared to negative.
% The intuition behind this is the following, in music often repetition tends to be salient on instants corresponding to beats and/or downbeat.
% \todo{remove some parts}
Given a set of hypotheses $\mathcal{H}$, our objective is to identify the underlying correct one/ones. 
To achieve this, we define for each $k$ a contrastive scoring function $h_k(\mathbf{x})$ based on audio features extracted from $\mathbf{x}$.
This function is designed to measure the mutual information between positive and negative elements chosen from $\mathcal{B}$ with respect to hypothesis $k$.
%$k$ designates the set of positive and negative elements.
The intuition is that, in music, repetition often becomes more salient at moments corresponding to beats and/or downbeats.
By leveraging this repetitive structure, the scoring function can effectively distinguish between hypotheses, prioritizing those that align with the rhythmic patterns inherent in the musical segment.

% \todo{add why we score like this}

% \remove[GP]{An hypothesis $\mathcal{H}_{\omega, \phi}= \{\omega, \phi\}$, define a candidate subset $\mathcal{B}_{\omega, \phi}$.} 

A hypothesis $\mathcal{H}_{\omega, \phi}$ defines a candidate subset $\mathcal{B}_{\omega, \phi}$.
From this subset, we sample an anchor time step $A_{\omega, \phi}$, followed by $n_p$ positive time steps $P_{\omega, \phi} = \{p^{i}_{\omega, \phi}\}_{i=1}^{n_p}$, selected from the remaining candidates in $\mathcal{B}_{\omega, \phi} \setminus A_{\omega, \phi}$.  
For the negative time steps $N_{\omega, \phi} = \{m^{i}_{\omega, \phi}\}_{i=1}^{n_n}$  we include 
i) easy negatives, \textit{i.e.}, time steps not aligned with \ac{PLP} peaks; 
ii) hard negatives, \textit{i.e.}, time steps that correspond to \ac{PLP} peaks but do not belong to $\mathcal{B}_{\omega, \phi}$.  
We apply this procedure to each hypothesis $\mathcal{H}_{\omega, \phi}$, to obtain its associated set of triplets $\mathcal{T}_{\omega, \phi} = \left\{ A_{\omega, \phi}, P_{\omega, \phi}, N_{\omega, \phi} \right\}$.
For conciseness, we denote by $\mathcal{H} = \left\{ \mathcal{H}_{k} \right\}_{k=1}^{k=K}$ the set of hypotheses and by $  \mathcal{T}_k = \left\{  A_k, P_k, N_k \right\}$ their associated triplets.
% \todo{better formulation}

% \begin{equation}
%   \mathcal{T}_k = \left\{  A_k, \mathbf{P}_k = \left\{ p_k^i \right\}_{i=1}^{n_p}, \mathbf{M}_k = \left\{ m_k^j \right\}_{j=1}^{n_n} \right\}
% \end{equation}

% \paragraph{Scoring}
% From the audio signal $ \mathbf{x} = \{x_l\}_{l=1}^{l=L}$, we extract the sequence of audio features $ \mathbf{F}^{\mathbf{x}} = [\,\mathbf{f}^{\mathbf{x}}_{1}, \mathbf{f}^{\mathbf{x}}_{2}, \dots, \mathbf{f}^{\mathbf{x}}_{T} ] \in \mathbb{R}^{d_f \times T}$ where $ \mathbf{f}^{\mathbf{x}}_{t} $ corresponds to the audio feature vector extracted at time frame $ t $.
% % \todo{better info here}
% For each hypothesis $ \mathcal{H}_{k} $ and its associated triplet $ \mathcal{T}_{k} $ we compute the contrastive scoring function $h_k(\mathbf{x})
%   = \mathcal{L}_{\text{NT-Xent}}(\bm{\mathrm{F}}^{\mathbf{x}}, \mathcal{T}_{k})$, using:

From the audio signal $\mathbf{x} = \{x_l\}_{l=1}^{L}$, we extract a sequence of feature vectors $\mathbf{V}^{\mathbf{x}} = [\,\mathbf{v}^{\mathbf{x}}_{1}, \mathbf{v}^{\mathbf{x}}_{2}, \dots, \mathbf{v}^{\mathbf{x}}_{T} ] \in \mathbb{R}^{d_v \times T}$, where $\mathbf{v}^{\mathbf{x}}_{t}$ is the audio feature corresponding to time frame $t$. 
For each hypothesis $\mathcal{H}_k$, we define the contrastive scoring function
\begin{equation}
  % \mathcal{L}^{\bm{\mathrm{F}}^{\mathbf{x}}}_{k}
  %h_k(\bm{x})
  %= 
h_k(\mathbf{x})\!=\!-\!\sum_{t_p \in P_{k}}\!\log\!
  \frac{\exp(\text{sim}(\mathbf{v}^{\mathbf{x}}_{A_{k}}, \mathbf{v}^{\mathbf{x}}_{t_p})/\tau)}{
  \sum_{t_n \in N_{k}} \exp(\text{sim}(\mathbf{v}^{\mathbf{x}}_{A_{k}}, \mathbf{v}^{\mathbf{x}}_{t_n})/\tau)},
\end{equation}
where $\tau$ is a temperature parameter and $\mathrm{sim}(\mathbf{a}, \mathbf{b})$ denotes cosine similarity: $\mathrm{sim}(\mathbf{a}, \mathbf{b}) \;=\; \frac{\mathbf{a}^\top \mathbf{b}}{\|\mathbf{a}\| \,\|\mathbf{b}\|}$.
  
\subsection{Training strategy}
Given the scores $h_k$ we extract a subset of select hypotheses with $\mathbf{s}$.
For each selected hypothesis, $\mathcal{H}_k \in \mathcal{H}^{\ast}$ we compute the following loss:
\begin{equation}
\begin{aligned}
  \mathcal{L}_{k}(\mathbf{z})
  &= \mathcal{L}_{\text{NT-Xent}}(\mathbf{z}, \mathcal{T}_{k}) \\
  &=\!-\!\sum_{t_p \in P_{k}}\!
  \log\!\frac{\exp(\text{sim}(\mathbf{z}^{k}_{A_{k}}, \mathbf{z}^{k}_{t_p})/\tau)}
  {\sum_{t_n \in N_{k}} \exp(\text{sim}(\mathbf{z}^{k}_{A_{k}}, \mathbf{z}^{k}_{t_n})/\tau)}.
\end{aligned}
\end{equation}

% Additionally, we investigated a \ac{WTA} strategy combined with metrical rhythmic levels.
% \todo{more detail here}

\section{Experiments}

\subsection{Datasets}

% \paragraph{Pretraining dataset} 
For the self-supervised \textit{pre-training} of our models, we used the Million Song Dataset \cite{Bertin-Mahieux2011}, a large collection of contemporary popular music tracks.
 We had access to audio excerpts  of approximately 905,000 tracks.
% We then extract quasi constant inter \ac{PLP} peaks  20s chunks (\ref{sec:details}). 
% We obtain such chunks spanning 600k tracks. We limit the number of chunks to 5 per tracks. 
% In total we had around 3M different chunks.
% Following explanations in \ref{sec:details} and \ref{sec:pretrain} we extract a fixed set of 20 seconds long audio chunks with quasi-constant inter-\ac{PLP} peaks, allowing a maximum of 5 chunks per track.
% These chunks were derived from approximately 600,000 pieces and resulted in a total of around 3 million unique chunks.
Following the explanations in Sections \ref{sec:details} and \ref{sec:pretrain}, we extracted a fixed set of 20-second audio chunks $\bf{x}$ from the tracks, allowing up to five chunks per track.
We only kept those with quasi-constant inter-\ac{PLP} peak intervals.
This ensures a controlled training set reducing \ac{PLP}-extraction related errors.
These chunks were selected from approximately 600,000 pieces, resulting in over 3 million unique chunks.

For \textit{fine-tuning} our models, we used the Ballroom \cite{ballroom,ballroom_2}, Beatles \cite{evaluation_beat}, Hainsworth \cite{hainsworth}, Harmonix \cite{oriol_nieto_2019_3527870}, HJDB \cite{HockmanDF12}, SMC \cite{smc} and RWC-Popular \cite{Goto2002RWCMD, rwc2} datasets.
In addition, we kept the GTZAN dataset \cite{GTZAN,GTZAN_2}, as commonly done in previous works, to serve as an unseen test set for evaluating generalization performance.
While additional annotated datasets exist, they are not openly accessible.\footnote{
While \cite{foscarin2024beatthis} provides pre-computed features as 22,050 Hz Log-Mel Spectrograms (LMS) for other datasets, their audio is not accessible.
We attempted to get the audio back from the LMS using the Griffin-Lim algorithm but the resulting quality was insufficient.
Therefore we chose not to include these datasets in our experiments, to avoid unfair comparisons.}

\subsection{Network Architecture}
\label{ap:arch}

The present work does not focus on designing a new deep network architecture.
We therefore adopt the openly available model from \cite{foscarin2024beatthis} as our backbone.
This model provides the current state-of-the-art results in beat and downbeat tracking.

Our pre-training model consists of  
$i)$ a frontend,  
$ii)$ an encoder network that outputs a sequence of representations, and  
$iii)$ our proposed projection layers $f_{\theta}^k$ for each hypothesis $k$.
The \textit{input} is a 128-bin mel spectrogram where the magnitudes are scaled by $\ln(1 + 1000x)$. 
The \textit{frontend}’s role is to integrate information across the 128 frequency bands into feature vectors.
It comprises three blocks, each containing two Partial Transformers inspired by \cite{Lu2023MusicSS}, which first process time, then frequency, followed by a 2D convolution, batch normalization, and GeLU activation.  
The first Partial Transformer treats each time step as a sequence across frequency bands, capturing harmonic structure, while the second operates across time steps for temporal modeling.  
Each block reduces the number of frequency bands while increasing channels, progressively refining the feature representation.  
A 2D convolution further processes the extracted features while preserving spatial structure.  
After three blocks, the output is reshaped and projected into a $T \times 512$ representation.

The \textit{encoder} network consists of six transformer layers with 16 heads of dimension 32, using rotary positional embeddings \cite{Su2021RoFormerET} and sigmoid gating \cite{NEURIPS2023_edbcb758}.  
The feed-forward dimension is set to 2048.  
The sequence is then processed by the hypothesis heads, which consist of $K$ linear layers that project the 512-dimensional sequence into a 64-dimensional sequence, followed by RMS normalization.  
In total, the network contains 20.3 million parameters. 

\begin{table*}[ht!]
    \centering
    \caption{Beat and Downbeat detection results with single split fine-tuning. Metrics are computed on the GTZAN dataset. $\star$ stand for the backbone from \cite{foscarin2024beatthis} and $\dagger$ the one used in \cite{gagnere:hal-04768296}}
    \vskip 0.15in
    \resizebox{\textwidth}{!}{ % Ensures the table fits within the column width
    \begin{small}
    \begin{sc}
    \begin{tabular}{llcccccccc}
        \toprule
         \multicolumn{1}{l}{pre-training} & \multicolumn{1}{l}{fine-tuning} & \multicolumn{1}{c}{back-} & \multicolumn{3}{c}{beat} & \multicolumn{3}{c}{downbeat} & \multirow{2}{*}{avg}\\
         \cmidrule{4-9} 
        \multicolumn{1}{l}{method} & \multicolumn{1}{l}{method} & \multicolumn{1}{c}{bone} & F1 & CMLt & AMLt & F1 & CMLt & AMLt & \\
        \midrule
        wta & bce-dbn & $\star$ & 88.7 & 81.0 & 92.4 & 75.8 & \underline{71.7} & \underline{87.5} & 82.9 \\
        wta & st-bce & $\star$ & \underline{89.2} & 79.8 & 90.4 & \underline{76.9} & 63.5 & 78.6  & 79.7\\
        2-wta & bce-dbn & $\star$ & 88.5 & 80.4 & 92.2 & 75.2 & 70.5 & 86.7 & 82.3\\
        2-wta & st-bce & $\star$ & 88.8 & 79.4 & 89.5 & 75.4 & 60.7 & 75.3 &  78.2\\
        3-wta & bce-dbn & $\star$ & \underline{89.2} & \underline{82.1} & \underline{92.8} & 75.3 & \underline{71.7} & \underline{87.5} & \underline{83.1}\\
        3-wta & st-bce & $\star$ & 88.9 & 79.9 & 89.7 & 75.9 & 61.9 & 77.6 & 79.0\\
        3-wta & bce-dbn & $\dagger$ & 88.1 & 80.6 & 91.6 & 75.7 & 71.5 & 87.0 & 82.4\\
        3-wta & st-bce & $\dagger$ & 88.0 & 78.5 & 88.6 & 73.9 & 58.3 & 74.0 & 76.9\\
        \midrule
        % Madmom Pseudo & bce-dbn & $\star$ & 89.5 & 82.0 & \textbf{93.2} & 77.2 & 73.5 & 87.5 &83.8\\
        % Madmom Pseudo & st-bce & $\star$ & \textbf{90.1} & 81.9 & 90.8 & 75.9 & 60.6 & 76.0 &79.2\\
        self-training & bce-dbn & $\star$ & 89.6 & \textbf{82.6} & 92.5 & \textbf{78.3} & \textbf{74.7} & \textbf{88.2} & \textbf{84.3}\\
        self-training & st-bce & $\star$ & 89.6 & 81.8 & 91.8 & 77.5 & 67.0 & 80.8 &81.4\\
        \midrule
        \multicolumn{3}{l}{\cite{Bck2020DeconstructAR}} & 88.5 & 81.3 & 93.1 & 67.2 & 64.0  & 83.2 & 79.6\\
        \multicolumn{3}{l}{\cite{Hung2022}} & 88.7 & 81.2 & 92.0 & 75.6 & 71.5 & 88.1 & 82.9\\
        \multicolumn{3}{l}{\cite{ZhaoXW22}} & 88.5 & 80.0 & 92.2 & 71.4 & 66.5 & 84.4 & 80.5 \\ 
        \multicolumn{3}{l}{\cite{foscarin2024beatthis}} & 89.1 $\pm$ 0.3 & 79.8 $\pm$ 0.6 & 89.8 $\pm$ 0.4 & \textbf{78.3} $\pm$ \textbf{0.4} & 67.3 $\pm$ 0.8 & 79.1 $\pm$ 0.6  &80.6\\
        \multicolumn{3}{l}{-- limited to data of \cite{Hung2022}} & 88.9 $\pm$ 0.1 & 79.9 $\pm$ 0.4 & 89.4 $\pm$ 0.2 & 75.5 $\pm$ 0.5 & 60.8 $\pm$ 1.2 & 75.5 $\pm$ 0.5 & 78.4\\
        
        \bottomrule
    \end{tabular}
        \end{sc}
    \end{small}
}
    \label{tab:gtzan}
\end{table*}

\subsection{Training Setup}
\label{sec:details}

The input, $\mathbf{x}$, is a 20-s long chunk of audio sampled at 16 kHz.
We turn it into a Mel-spectrogram, which is computed with a hop size of 20 ms (320 samples);  this is therefore the time granularity of our representation. 
For \ac{PLP} extraction, we first compute the \ac{OSF} using the Superflux method \cite{superflux} and the same hop length (this is because the \ac{PLP} and input time frames should be aligned), followed by a high-pass filter.
% The \ac{PLP} is then obtained using Librosa's implementation \cite{mcfee2015librosa}, and peak picking is performed with Scipy's \texttt{find\_peaks} function.
% The onset strength function is computed using the Superflux method \cite{superflux}, followed by a high-pass filter.
% \todo{add hop size parameters}
% The \ac{PLP} is computed with Librosa implementation \cite{mcfee2015librosa}, and peak picking is performed using Scipy's \texttt{find\_peaks} function.

% We work with audio sampled at 16 Khz.
% We compute the onset strength function with the Superflux \cite{superflux} which is then high-pass filter.
% \todo{add hop size parameters}
% We used Librosa for the PLP computation.
% Finally, we applied peak picking with Scipy's \texttt{find\_peaks} function.
We discard chunks where the distance between successive \ac{PLP} peaks is not approximately constant.
These variations may result from actual changes in the music (such as tempo shifts, the introduction of new elements, or silences), or from inconsistencies in the \ac{PLP}. 
Such cases typically indicate a phase and/or ratio shift, as illustrated in Fig. \ref{fig:plp}, where the ratio shifts from 2 to 3. 
When this occurs, it becomes impossible to determine a fixed ratio and phase to explain the beat sequence from the \ac{PLP} peaks, violating the assumption in equation \ref{eq:subseq}.
In practice, we discard chunks if the relative variation in inter-peak distances exceeds 20\% between two successive peaks.
This threshold is high enough to allow for slight tempo variations but prevents phase and/or ratio shift to occur.

% \todo{add in appendix chunks accuracies with annotated data}

Based on preliminary studies on annotated data, we restrict the set of considered ratios to $\Omega = \left\{ 1, 2, 3, 4\right\}$ resulting in a total of $K = \sum_{\omega \in \Omega} \omega$ = 10 hypotheses.
Indeed the \ac{PLP} rarely detects rhythmic events going faster than four times the beat. 
Our scoring function $h$ is computed for each hypothesis with Variable-Q chromagram features also computed with a 20ms hop size. 
We found that it was beneficial to stack input features with time-lagged copies of itself. 
% \todo{More explanations here}

% \subsection{Pre-training Details}
% \label{sec:pretrain}

To improve scoring stability, we trained models on a fixed set of chunks $\bf{x} \in \mathcal{X}$.
At each epoch $e$, we store the scores  $h^e_k(\bf{x})$ for each chunk $\bf{x}$  and each hypothesis $\mathcal{H}_k \in \mathcal{H}$.
These scores depend on the random sampling of the anchor, positives and negatives of $\mathcal{T}_k$ (\ref{sec:sampling}). 
We therefore use the running mean over all past epochs $h^{[0:e]}_k(\bf{x})$ as the final score for hypothesis selection.
This aggregation reduces score fluctuations, ensuring a more stable selection function and a smoother training process.

Our models were pre-trained for 200,000 steps in FP16 precision using 8 V100 GPUs with a global batch size of 128.
Gradient clipping was applied to stabilize training.
We used AdamW optimizer \cite{loshchilov2018decoupled} alongside a cosine learning rate scheduler.
The learning rate was linearly increased to 0.0005 over the first 32,000 steps and then gradually decreased for the remaining training steps.

% To improve scoring stability, we trained models on a fixed set of chunks. At each epoch we store, for each chunk, the obtained scores $h_k$ for all hypotheses $H_k$.
% The score is computed as a running average over all previous epochs, including the current one.
% This aggregation smooths score variations, leading to a more stable selection function and training process.
% \todo{maybe add algorithm for chunks and stuff}

% After \ac{ssl} pre-training, we need to adapt the model to the downstream task of beat tracking.
% This is done by adding a linear classification probe $g(.)$ and fine-tuning both the encoder and the linear probe.
% $g(.)$ projects the embedding into the scalar beat activation function.
% Instead of feeding $g(.)$ with the output of the encoder, we feed it with a weighted sum of the outputs of each layer of the Transformer~\cite{Yang2021SUPERBSP}.
% That is $z = \sum_{l=1}^{8} \alpha_l z^{(l)}$, where $z^{(l)}$ is the output of layer $l$.
% The weights $\alpha_l$ are jointly learned with the linear probe $g(.)$.

% The system is trained to minimize the binary cross-entropy loss between the beat activations and the target.
% Following the literature we widened the beat targets by a window [0.25, 0.5, 1, 0.5,0.25] \cite{Bck2020DeconstructAR}.
% We used the Adam optimizer \cite{adam} with an initial learning rate of 1e-5 and a polynomial decay learning rate scheduler.

% The \ac{dbn} is configured to model a tempo range of 40-270  beats per minute with transition lambda set to 45, observation lambda to 9, and a threshold of 0.15.

\subsection{Fine-tuning methods}

After pre-training we discard our ensemble of heads $f_\theta$ and use directly the intermediate representations $\mathbf{z}_t$.
These are projected, with a linear layer, into two scalar outputs that represent beat and downbeat activations respectively. 
We explored two different fine-tuning methods. 

\textbf{BCE-DBN:}
The first method trains these activations to minimize the \ac{BCE} loss, with a target-widening strategy, the neighboring frames of  a frame annotated as positive are also set as positive targets but with a lower weight of 0.5 \cite{Bck2020DeconstructAR}. 
Once training is complete, the activations are post-processed using a \ac{DBN} \cite{florian_krebs_2018_1414966} to produce the final beat and downbeat positions.

\textbf{ST-BCE:}
The second approach \cite{foscarin2024beatthis} introduces a Shift-Tolerant Weighted BCE (ST-BCE) loss function to address 
i) class imbalance, 
ii) annotations' imprecision and 
iii) the reliance on a \ac{DBN} for post-processing.
In the ST-BCE approach, positive examples are weighted by a factor $\alpha$ defined as the ratio of negative to positive frames, computed separately for beats and downbeats.
Predictions are max-pooled over time before comparing them to ground-truth labels, ensuring that the strongest positive activation within a local window is considered, while negative labels near annotated beats are ignored. This reduces sensitivity to slight misalignment and prevents over-penalization of near-correct predictions. Formally the loss is defined as:
$L_{\text{st}}(\mathbf{y}, \hat{\mathbf{y}}) = -\sum_t \alpha y_t \log\big(m_7(\hat{\mathbf{y}})_t\big) + (1 - m_{13}(\mathbf{y})_t) \log\big(1 - m_7(\hat{\mathbf{y}})_t\big)$ where $m_l$ denotes max-pooling over $l$ frames. Beat and downbeat detection are obtained by applying peak-picking on their respective activations functions.  

% \todo{We add a projection head}
In each setting, we trained both the \textit{front-end} and \textit{encoder} for 100 epochs using the AdamW optimizer \cite{loshchilov2018decoupled} with a batch-size of 64.
During the initial 3 epochs, the \textit{encoder} weights were kept frozen.
The learning rate was scheduled with a cosine annealing strategy, ramping up to 0.001 over the first 400 optimization steps and gradually decaying to 0 by the final epoch.
We employ the same data augmentations as those of Beat This \cite{foscarin2024beatthis}.
Every track can be either pitch-shifted by a transposition factor sampled in [-5, +6] semitones or time-stretched with a factor in 20, 16, 12, 8, and 4\% both faster and slower.
% Additionally, we also use their masking strategy. \todo{add explanation here}

\begin{table*}[ht]
    \centering
    \caption{Average cross-validation scores for Beat (B) and Downbeat (DB) detection systems on reference datasets.}
    \vskip 0.15in
    \resizebox{\textwidth}{!}{
    % \begin{center}
    \begin{small}
    \begin{sc}
    \begin{tabular}{llcccccccccccc}
    \toprule
    \multicolumn{1}{l}{pre-training} & \multicolumn{1}{l}{fine-tuning} & \multicolumn{2}{c}{ballroom} & \multicolumn{2}{c}{beatles} & \multicolumn{2}{c}{harmonix} & \multicolumn{2}{c}{hainsworth} & \multicolumn{2}{c}{rwc popular} & \multicolumn{1}{c}{smc} & \multirow{2}{*}{avg} \\
    \cmidrule{3-13} 
    \multicolumn{1}{l}{method} & \multicolumn{1}{l}{method} & b & db & b & db & b & db & b & db & b & db & b & \\
    \midrule
    wta & bce-dbn & 96.8 & \underline{95.2} & 90.8 & 82.4 & 93.0 & 88.6 & 89.8 & 79.4 & 92.1 & 91.8 & \underline{58.3} & 87.1 \\
    wta & st-bce & \underline{96.9} & 93.6 & 90.9 & 77.7 & 93.2 & 85.0 & 89.0 & 72.2 & 91.3 & 88.8 & 54.7 & 84.8 \\
    3-wta & bce-dbn & 96.8 & 95.0 & \underline{91.3} & \underline{83.7} & \underline{93.4} & \underline{89.1} & 89.0 & \underline{79.8} & 92.4 & \underline{92.4} & 58.2 & \underline{87.4} \\
    3-wta & st-bce & 96.6 & 93.4 & 90.8 & 78.2 & 93.1 & 84.9 & \underline{90.3} & 72.3 & \underline{92.6} & 90.6 & 56.1 & 85.4 \\
    \midrule
    % Madmom pseudo lab + bce-dbn & 97.3 & 95.9 & 91.7 & 83.7 & 93.2 & 89.5 & 90.0 & 80.3 & 91.5 & 91.1 & 58.0 & 87.5 \\
    % Madmom pseudo lab + st-bce & 97.3 & 94.5 & 92.0 & 80.0 & 93.4 & 85.9 & 91.0 & 73.5 & 91.2 & 87.7 & 56.7 & 85.7 \\
    self-training & bce-dbn & 97.5 & \textbf{97.2} & 93.6 & \textbf{86.0} & 95.0 & \textbf{92.4} & \textbf{93.2} & \textbf{83.6} & 91.3 & 91.0 & \textbf{61.9} & \textbf{89.3} \\
    self-training & st-bce & \textbf{97.8} & 95.7 & \textbf{94.2} & 82.0 & \textbf{95.3} & 88.4 & \textbf{93.2} & 77.3 & 91.1 & 89.3 & 60.3 & 87.7 \\
    \midrule
    \multicolumn{2}{l}{\cite{foscarin2024beatthis}}  & 97.0 & 94.1 & 91.6 & 81.5 & 93.2 & 85.9 & 88.9 & 72.9 & 93.3 & 90.0 & 58.4 & 86.1 \\
    \multicolumn{2}{l}{\cite{Bck2020DeconstructAR}} & 95.7 & 93.0 & 81.4 & - & 90.4 & 80.8 & 89.7 & 76.3 & - & - & 55.3 & 60.2 \\
    \multicolumn{2}{l}{\cite{Hung2022}} & 95.6 & 94.4 & 92.6 & 84.9 & 95.0 & 90.3 & 88.5 & 78.5 & \textbf{94.4} & \textbf{94.8} & 59.4 & 88.0 \\
    \multicolumn{2}{l}{\cite{ZhaoXW22}} & 96.3 & 95.1 & - & - & 93.9 & 89.3 & 88.7 & 76.7 & - & - & 56.2 & 54.2 \\
    \bottomrule
    \end{tabular}
    \end{sc}
    \end{small}
    % \end{center}
    }
    \label{tab:crossval}
\end{table*}
\subsection{Evaluation metrics}

We report the standard evaluation metrics commonly used in the literature \cite{evaluation_beat, Davies2014EvaluatingTE}.
The F-measure considers a detected beat correct if it falls within a ±70-ms tolerance window around a ground-truth beat position.
Additionally, we use continuity-based metrics, where a beat is valid only if the previous beat is also correct.
The CMLt metric measures the proportion of beats correctly aligned with annotations at the expected metrical level under this constraint.
The AMLt metric extends this by allowing metrical variations, such as double or half tempos or off-beat shifts.
We computed these metrics using the \texttt{mir\_eval} package \cite{raffel2014mir_eval}, following the convention of trimming beats occurring in the first 5 seconds and using default parameters.

\subsection{Single-split fine-tuning}
% \todo{Add interval in table}

We split each dataset (except GTZAN that we keep for evaluation) into train and validation (7/8, 1/8) and use the union of the train splits to fine-tune our pre-trained models.
We use the same splits as \cite{foscarin2024beatthis}. 
This serves as an experiment to explore the behavior of the sampling mechanism $s$.
We explored 3 different mechanisms: \ac{WTA}, 2-\ac{WTA} and 3-\ac{WTA}.
We also compare the backbone architecture of \cite{foscarin2024beatthis}, to the one used in \cite{gagnere:hal-04768296}. The latter consists of a mel-spectrogram fed to a transformer encoder without rotary positional encoding.

We report the results in the upper part of Table~\ref{tab:gtzan}.
The lower part of the Table provides baselines from state-of-the-art systems.
It is important that, at the exception of \cite{foscarin2024beatthis} the performances are not obtained in the same conditions.

We see that the backbone of \cite{foscarin2024beatthis} consistently outperforms \cite{gagnere:hal-04768296} under equivalent pre-training and fine-tuning configurations.

Using 3-WTA improves beat metrics compared to one- or two-WTA variants: 3-WTA + BCE-DBN achieves the highest Beat F1 (89.2), CMLt (82.1), and AMLt (92.8).
Furthermore, a trade-off emerges, between using the BCE with \ac{DBN} and ST-BCE with peak-picking echoing prior work~\cite{foscarin2024beatthis}.
While BCE-DBN generally leads to superior continuity metrics (CMLt, AMLt), ST-BCE can lead to better F1.
This trade-off is also evident in the downbeat results, where ST-BCE yields higher F1 (e.g., 76.9 for 1-WTA).

\subsection{Cross-validation}

In the second experiment, we split each dataset into train, validation and test (6/8, 1/8, 1/8), use the union of the trains to fine-tune our pre-trained models and the union of the tests for evaluation.
The folds are changed following and 8-fold cross-validation.
This is the methodology commonly used in the literature.
Here also we used the same folds of \cite{foscarin2024beatthis}.

The upper part of Table~\ref{tab:crossval} presents the average beat and downbeat metrics for our pre-trained models, while the lower part shows the performance of existing systems. Additionally, Tables \ref{tab:cv_beat} and \ref{tab:cv_downbeat} in Appendix \ref{sec:full_metrics} provide detailed metrics for beat and downbeat, respectively. Cross-validation results are reported only for WTA and 3-WTA pre-training, as they demonstrated the best performance.

Using 3-WTA for pre-training leads to improved average performance across both fine-tuning strategies.
Specifically, 3-WTA + BCE-DBN achieves an 87.4 average, slightly exceeding the 87.1 from WTA + BCE-DBN.
We also observe particularly strong performance on Hainsworth (Beat: 89.0, Downbeat: 79.8).
However, performance on RWC Popular (92.4 for both beat and downbeat) lags behind that of \cite{Hung2022}, which consequently lowers our overall average compared to their 88.0.
Meanwhile, the 3-WTA + ST-BCE setup (85.4 average) is slightly below \cite{foscarin2024beatthis} (86.1); however, our fine-tuning is performed on a smaller training set (3144 vs. 4556 tracks), which likely accounts for the difference.

\subsection{Self-Training}

% The previous results were obtained using a multi-hypothesis contrastive pre-training approach.
% To explore than upper bound on performance, we aim to estimate the best-case scenario, where the correct hypothesis is always selected during training. However, in the context of self-supervised learning (SSL), the ground-truth beat positions are not available, making it infeasible to directly identify the correct hypothesis.

% To approximate this ideal condition, we employ a self-training strategy based on pseudo-labeling. 
% Specifically, pseudo-labels are generated to serve as proxy beat annotations, enabling the selection of the most relevant hypothesis, this is equivalent to the above \ac{WTA} pre-training where the scoring function $s$ always detect the right underlying hypothesis. 
% Given the pseudo beat positions, the same sampling procedure described in Section~\ref{sec:sampling} is applied, constrained to the case $\omega=1$.
% We don't use the ensemble of hypotheses heads.
% Subsequently, the contrastive loss is computed directly on the sequence of learned representations, $\bm{z}_t$.

The previous results were obtained using the proposed multi-hypothesis contrastive pre-training approach. To establish an upper bound on performance, we investigate an advanced setting in which the correct hypothesis is consistently selected during training. However, in the context \ac{SSL}, ground-truth annotations are not available, making it infeasible to directly determine the correct hypothesis.

To approximate this ideal condition, we employ a self-training strategy based on pseudo-labeling.
Specifically, we generate pseudo-labels as surrogate beat annotations, which enables the selection of the most relevant hypothesis.
This approach is equivalent to the \ac{WTA} pre-training paradigm, where the scoring function $s$ always identifies the correct underlying hypothesis. Given these pseudo beat positions, we apply the same sampling procedure described in Section~\ref{sec:sampling}, with $\omega=1$.
Unlike the training method described in \ref{sec:gen}, we do not use the ensemble of hypothesis-specific projection heads $f_\theta$.
Instead, the contrastive loss is computed directly on the general sequence of learned representations, $\bf{z}_t$.
% To evaluate the effectiveness of this approach, we compare different pseudo-labeling strategies and their impact on model performance. % Further details on the pseudo-labeling procedure will be provided in Section~\todo{reference section}.

We report the system accuracy in single split (middle part of Table \ref{tab:gtzan}) and averaged metric for cross-validation fine-tuning (middle part of Table \ref{tab:crossval}). Detailed metrics for 8-fold cross-validation are gathered in \ref{sec:full_metrics}.

In Table~\ref{tab:gtzan}, self-training + BCE-DBN achieves an 84.3 average on GTZAN, outperforming all other listed methods.
This gain comes from notable improvements in both beat detection (+0.5 F1 and +1.4 CMLt) and downbeat detection (+2.7 F1 and +3.2 CMLt).

In Table~\ref{tab:crossval}, the same approach reaches the highest average score 89.3 — which is +1.3 above \cite{Hung2022} and +3.2 above \cite{foscarin2024beatthis}. The largest gains appear on Hainsworth (+3.5 average beat metrics, +5.1 average downbeat metrics) and SMC (+2.5), a particularly challenging dataset. RWC Popular sees lower performance (-3.1 and -3.8), but on the remaining datasets both fine-tuning methods match or surpass prior state-of-the-art methods.
Overall, these results confirm that self-training provides a robust performance boost for beat and downbeat detection.

\section{Conclusion}
We proposed Knowledge Driven Multi Hypothesis Learning to guide contrastive \ac{SSL}.
In this framework we rely on domain knowledge to score, at each step and for each sample, a set of hypotheses, select the n-winning ones, and use those for training.
We instantiated this framework for the task of rhythm analysis and explored different hypothesis selection mechanisms for pre-training. 
We fine-tuned the model with annotated data and demonstrated state-of-the-art performance on several benchmarking datasets, confirming the effectiveness of our pre-training.
Additionally, we explored self-training as an advanced setting of our framework. 
Pre-training a model this way yielded state-of-the-art performance on most benchmarks, often surpassing previous systems by 2\%. 
Future work will focus on using a more diverse set of datasets for pre-training and exploring alternative selection mechanisms incorporating musical meter knowledge.

\newpage
\bibliography{bib}
\bibliographystyle{icml2025}

%%%%%%%%%%%%%%%%%%%%%%%%%%%%%%%%%%%%%%%%%%%%%%%%%%%%%%%%%%%%%%%%%%%%%%%%%%%%%%%
%%%%%%%%%%%%%%%%%%%%%%%%%%%%%%%%%%%%%%%%%%%%%%%%%%%%%%%%%%%%%%%%%%%%%%%%%%%%%%%
% APPENDIX
%%%%%%%%%%%%%%%%%%%%%%%%%%%%%%%%%%%%%%%%%%%%%%%%%%%%%%%%%%%%%%%%%%%%%%%%%%%%%%%
%%%%%%%%%%%%%%%%%%%%%%%%%%%%%%%%%%%%%%%%%%%%%%%%%%%%%%%%%%%%%%%%%%%%%%%%%%%%%%%
\newpage
\appendix
\onecolumn

\section{Preliminary study}

We evaluated the ability to detect the underlying hypotheses with different features. 
From an audio $\mathbf{x} = \{x_l\}_{l=1}^{L}$ we extract a sequence of features vector $\mathbf{V}^{\mathbf{x}} = [\,\mathbf{v}^{\mathbf{x}}_{1}, \mathbf{v}^{\mathbf{x}}_{2}, \dots, \mathbf{v}^{\mathbf{x}}_{T} ] \in \mathbb{R}^{d_v \times T}$.
The scoring function for a triplet $\mathcal{T}_{k} = \{ A_k, \bf{P}_k, \bf{N}_k \}$ is defined as follows:

\begin{equation}
  % \mathcal{L}^{\bm{\mathrm{F}}^{\mathbf{x}}}_{k}
  %h_k(\bm{x})
  %= 
h_k(\mathbf{x})\!=\!-\!\sum_{t_p \in \bf{P}_{k}}\!\log\!
  \frac{\exp(\text{sim}(\mathbf{v}^{\mathbf{x}}_{A_{k}}, \mathbf{v}^{\mathbf{x}}_{t_p})/\tau)}{
  \sum_{t_n \in \bf{N}_{k}} \exp(\text{sim}(\mathbf{v}^{\mathbf{x}}_{A_{k}}, \mathbf{v}^{\mathbf{x}}_{t_n})/\tau)},
\end{equation}
where $\tau$ is a temperature parameter and $\mathrm{sim}(\mathbf{a}, \mathbf{b})$ denotes cosine similarity: $\mathrm{sim}(\mathbf{a}, \mathbf{b}) \;=\; \frac{\mathbf{a}^\top \mathbf{b}}{\|\mathbf{a}\| \,\|\mathbf{b}\|}$.

In a preliminary study, we explored the accuracy of our scoring mechanism. 
Given annotated data we a fixed set of 20s long chunks.
With ground truth annotations, we know the correct underlying sequence of \ac{PLP} peaks corresponding to beats.
The exact accuracy metrics correspond when we detect correct hypothese. 
The octave one when we detect a metric that is acceptable.
As an example let's say the correct is $\omega=2$ and $\phi=1$.
Then when accounting for octave error $\omega=4$ and $\phi=1$ or $\phi=3$ are also deemed correct.
Finally the Metric level one account that a detection is correct when ratio $\omega$ is the same as the correct ratios. 

Table. \ref{tab:prelim} gathers the different accuracy for mel spectrum, Chroma features computed either from Variable-Q-Transform (VQT) or Short Time Fourier Transform (STFT) and Mel-frequency Cepstral Coefficients (MFCC). 
For each of these feature we report the top $k$ accuracies, $k \in \{1,2,3\}$ on three annotated datasets: Ballroom, GTZAN and SMC.
% We used $\tau=0.5$.

\begin{table*}[ht]
  \caption{Comparison of selection mechanism accuracy (in \%) with different audio features.
  % Metrics are reported on Ballroom, GTZAN and SMC datasets.
  }
  \centering
  \vskip 0.15in
  \resizebox{\textwidth}{!}{ 
  \setlength{\tabcolsep}{4pt}
  \scriptsize
  \begin{sc}
  \begin{tabular}{lccccccccccc}
      \toprule
      \multirow{2}{*}{Top K Accuracy} & \multirow{2}{*}{Feature} & \multicolumn{3}{c}{Ballroom} & \multicolumn{3}{c}{GTZAN} & \multicolumn{3}{c}{SMC} \\
      \cmidrule(lr){3-5} \cmidrule(lr){6-8} \cmidrule(lr){9-11}
      & & Exact & Octave & Metric Lvl & Exact & Octave & Metric Lvl & Exact & Octave & Metric Lvl \\
      \midrule
      \multirow{4}{*}{1}              & Mel & 19.5 & 82.7 & 20.7 & 20.9 & 85.2 & 22.8 & 20.2 & 62.8 & \textbf{31.8} \\
                                      & chroma stft & 14.3 & \textbf{85.3} & 14.8 & 17.3 & 88.1 & 18.4 & 14.0 & \textbf{66.7} & 22.5 \\
                                      & chroma vqt & \textbf{22.0} & 82.3 & \textbf{22.5} & \textbf{23.8} & 85.5 & \textbf{24.8} & \textbf{24.8} & 65.9 & 31.0 \\
                                      & mfcc & 6.2 & 71.5 & 10.1 & 8.4 & \textbf{88.2} & 12.0 & 5.3 & 60.5 & 11.2 \\

      \midrule
      \multirow{4}{*}{2}              & mel & 37.5 & 92.9 & 40.6 & 34.9 & 95.1 & 38.2 & 31.0 & \underline{80.6} & 44.2 \\
                                      & chroma stft & 46.3 & 93.1 & 48.5 & \underline{53.6} & \underline{95.2} & \underline{55.6} & 45.0 & 79.8 & 53.5 \\
                                      & chroma vqt & \underline{53.0} & \underline{94.8} & \underline{54.7} & 53.3 & 95.0 & \underline{55.6} & \underline{46.5} & \underline{80.6} & \underline{54.3} \\
                                      & mfcc & 21.1 & 84.0 & 34.1 & 32.6 & 92.3 & 44.9 & 22.1 & 78.4 & 41.9 \\

      \midrule
      \multirow{3}{*}{3}              & mel & 50.2 & 96.4 & 56.2 & 55.2 & \textbf{97.3} & 62.2 & 44.2 & 86.0 & 59.7 \\
                                      & chroma stft & 63.8 & 95.8 & 68.1 & \textbf{73.0} & 97.2 & \textbf{77.1} & \textbf{64.3} & 85.3 & \textbf{78.3} \\
                                      & chroma vqt & \textbf{69.6} & \textbf{96.7} & \textbf{72.7} & 69.3 & \textbf{97.3} & 73.0 & 59.7 & 86.8 & 66.7 \\
                                      & mel & 35.5 & 90.1 & 55.1 & 48.2 & 94.9 & 68.6 & 37.5 & \textbf{88.2} & 59.1 \\
      \bottomrule
  \end{tabular}
  \end{sc}}
  \label{tab:prelim}
\end{table*}

% \newpage
\section{Cross Validation Metrics}
\label{sec:full_metrics}

We report in Table \ref{tab:cv_beat} and Table \ref{tab:cv_downbeat} the full cross-validation scores for beat and downbeat, respectively.

\begin{table}[t]
\caption{Beat cross-validation scores. The best performances are shown in bold, and we have underlined the best-performing method for our proposed pre-training scheme.}
\vskip 0.15in
    \resizebox{\textwidth}{!}{ 
    \setlength{\tabcolsep}{4pt}
    \scriptsize
    % \begin{center}
    % \begin{small}
    \begin{sc}
    \begin{tabular}{ll|ccc|ccc|ccc}
    \toprule
     pre-training & fine-tuning & F1 & CMLt & AMLt & F1 & CMLt & AMLt & F1 & CMLt & AMLt\\
\midrule

% [0.2cm]
% ~ & ~ & ~ & ~ & ~ & ~ & ~ & ~ & ~ & ~ & ~ \\
% [0.1cm]
& & \multicolumn{3}{c|}{\rule{0pt}{0.3cm}ballroom} & \multicolumn{3}{c|}{\rule{0pt}{0.3cm}beatles} & \multicolumn{3}{c}{\rule{0pt}{0.2cm}harmonix} \\
[0.2cm]

 wta  & bce-dbn & 97.2 & 95.7 & \underline{97.4} & 94.0 & 87.0 & 91.4 & 95.4 &89.2 &94.3\\
 
 wta  & st-bce & \underline{97.5} & \underline{96.1} & 97.1 & \underline{94.6} & 87.4 & 90.8 & 95.9 &89.7 &94.1\\
 
 3-wta & bce-dbn & 97.2 & 95.8 & 97.3 & 94.5 & \underline{87.8} & \underline{91.7} & 95.6 &\underline{90.1} &\underline{94.5}\\
 
 3-wta & st-bce & 97.3 & 95.7 & 96.7 & 94.4 & 86.8 & 91.1 & \underline{96.0} &89.8 &93.5\\
% [0.3cm]
\midrule
 self-training & bce-dbn & 97.9 & 96.9 & 97.7 & 96.2 & 91.7 & 92.9 & 96.9 &92.5 &95.6\\
 self-training & st-bce & \textbf{98.2} & \textbf{97.4} & \textbf{97.8} & \textbf{96.4} & \textbf{92.1} & \textbf{94.1} & \textbf{97.3} &93.1 &95.6\\
% [0.3cm]
\midrule
 \multicolumn{2}{l|}{\cite{foscarin2024beatthis}} & 97.5 & 96.4 & 97.0 & 94.5 & 87.2 & 93.0 & 95.8 &89.9 &94.0\\
 \multicolumn{2}{l|}{\cite{Bck2020DeconstructAR}} & 96.2 & 94.7 & 96.1 & 83.7 & 74.2 & 86.2 & 93.3 &84.1 &93.8\\
 \multicolumn{2}{l|}{\cite{Hung2022}} & 96.2 & 93.9 & 96.7 & 94.3 & 89.6 & 93.8 & 95.3 &\textbf{93.9} &\textbf{95.9}\\
 \multicolumn{2}{l|}{\cite{ZhaoXW22}} & 96.8 & 95.4 & 96.6 & 0.0 & 0.0 & 0.0 & 95.4 &90.5 &95.7\\
[0.2cm]

& & \multicolumn{3}{c|}{hainsworth} & \multicolumn{3}{c|}{rwc} & \multicolumn{3}{c}{smc} \\
[0.2cm]

 wta & bce-dbn & 91.4 & 85.7 & 92.4 & 93.9 & 87.9 & \underline{94.6} & 60.1 &49.8 &\underline{65.0}\\
 wta & st-bce & 91.5 & 84.6 & 90.9 & 93.9 & 87.2 & 92.8 & 61.7 &46.8 &55.5\\
 3-wta & bce-dbn & 90.6 & 83.9 & \underline{92.5} & 94.3 & 88.6 & 94.4 & 59.8 &\underline{50.2} &64.7\\
 3-wta & st-bce & \underline{92.8} & \underline{87.1} & 91.0 & \underline{94.4} & \underline{89.2} & 94.3 & \underline{62.1} &48.0 &58.1\\
 % [0.3cm]
 \midrule
 self-training & bce-dbn & 94.3 & \textbf{90.8} & \textbf{94.6} & 93.7 & 86.8 & 93.3 & 62.9 &\textbf{54.6} &\textbf{68.1}\\
 self-training & st-bce & \textbf{94.8} & 90.4 & 94.5 & 93.5 & 86.0 & 93.7 & \textbf{64.2} &52.9 &63.7\\
 % [0.3cm]
\midrule
 \multicolumn{2}{l|}{\cite{foscarin2024beatthis}} & 91.9 & 84.0 & 90.9 & \textbf{96.1} & 90.1 & 93.6 & 62.7 &51.4 &61.0\\
 \multicolumn{2}{l|}{\cite{Bck2020DeconstructAR}} & 90.4 & 85.1 & 93.7 & 0.0 & 0.0 & 0.0 & 55.2 &46.5 &64.3\\
 \multicolumn{2}{l|}{\cite{Hung2022}} & 87.7 & 86.2 & 91.5 & 95.0 & \textbf{92.5} & \textbf{95.8} & 60.5 &51.4 &66.3\\
 \multicolumn{2}{l|}{\cite{ZhaoXW22}} & 90.2 & 84.2 & 91.8 & 0.0 & 0.0 & 0.0 & 59.6 &45.6 &63.5\\

    \bottomrule
    \end{tabular}
    \end{sc}
    % \end{small}
    % \end{center}
    }
\vskip -0.1in
\label{tab:cv_beat}
\end{table}

\begin{table}[t]
\caption{Downbeat cross-validation scores}
\vskip 0.15in
\resizebox{\textwidth}{!}{
  \setlength{\tabcolsep}{4pt}
  \scriptsize
  \begin{sc}
  \begin{tabular}{ll|ccc|ccc|ccc}
    \toprule
     pre-training & fine-tuning & F1 & CMLt & AMLt & F1 & CMLt & AMLt & F1 & CMLt & AMLt\\
\midrule

% [0.2cm]
% ~ & ~ & ~ & ~ & ~ & ~ & ~ & ~ & ~ & ~ & ~ \\
% [0.1cm]
& & \multicolumn{3}{c|}{\rule{0pt}{0.3cm}ballroom} & \multicolumn{3}{c|}{\rule{0pt}{0.3cm}beatles} & \multicolumn{3}{c}{\rule{0pt}{0.2cm}harmonix} \\
[0.2cm]

    wta           & bce-dbn       & 94.5 & \underline{93.9} & \underline{97.2}
                                      & 85.9 & 76.3             & 85.1
                                      & 90.2 & 84.7             & 90.8 \\
    wta           & st-bce        & 94.8 & 91.6             & 94.5
                                      & 86.6 & 67.9             & 78.6
                                      & 90.3 & 79.7             & 85.0 \\
    3-wta         & bce-dbn       & 94.2 & 93.5             & \underline{97.2}
                                      & \underline{87.4} & \underline{77.3} & \underline{86.4}
                                      & \underline{90.5} & \underline{85.7} & \underline{91.2} \\
    3-wta         & st-bce        & \underline{94.9} & 91.5       & 93.9
                                      & 86.7 & 67.7             & 80.3
                                      & 90.4 & 79.7             & 84.5 \\
    \midrule

    self-training & bce-dbn       & \textbf{97.0} & \textbf{96.6} & \textbf{97.9}
                                      & \textbf{89.6} & \textbf{81.4} & \textbf{87.1}
                                      & \textbf{93.9} & \textbf{90.3} & \textbf{93.1} \\
    self-training & st-bce        & 96.6         & 94.3         & 96.1
                                      & 89.4         & 74.4         & 82.2
                                      & 93.1         & 84.5         & 87.7 \\
    \midrule

    \multicolumn{2}{l|}{\cite{foscarin2024beatthis}}    & 95.3 & 92.9 & 94.1
                                                     & 88.8 & 73.8 & 82.0
                                                     & 90.7 & 81.2 & 85.9 \\
    \multicolumn{2}{l|}{\cite{Bck2020DeconstructAR}}   & 91.6 & 91.3 & 96.0
                                                     &  0.0 &  0.0 &  0.0
                                                     & 80.4 & 74.7 & 87.3 \\
    \multicolumn{2}{l|}{\cite{Hung2022}}                      & 93.7 & 92.7 & 96.8
                                                     & 87.0 & 81.2 & 86.5
                                                     & 90.8 & 87.2 & 92.8 \\
    \multicolumn{2}{l|}{\cite{ZhaoXW22}}               & 94.1 & 94.4 & 96.9
                                                     &  0.0 &  0.0 &  0.0
                                                     & 89.8 & 86.3 & 91.9 \\
    [0.2cm]

    & & \multicolumn{3}{c|}{Hainsworth}
      & \multicolumn{3}{c|}{rwc}
      & \multicolumn{3}{c}{SMC} \\
    [0.2cm]

    wta           & bce-dbn       & 78.6 & \underline{73.4} & 86.2
                                      & 92.7 & 89.1             & 93.6
                                      &   –  &    –               &   –   \\
    wta           & st-bce        & 79.2 & 63.0             & 74.3
                                      & 91.7 & 85.9             & 88.7
                                      &   –  &    –               &   –   \\
    3-wta         & bce-dbn       & 78.9 & 73.1             & \underline{87.3}
                                      & \underline{93.2} & \underline{89.8} & \underline{94.1}
                                      &   –  &    –               &   –   \\
    3-wta         & st-bce        & \underline{80.1} & 62.1       & 74.6
                                      & 92.9             & 88.1             & 90.9
                                      &   –  &    –               &   –   \\
    \midrule

    self-training & bce-dbn       & 83.2 & \textbf{78.6} & \textbf{88.9}
                                      & 92.4 & 87.8         & 92.7
                                      &   –  &    –         &   –   \\
    self-training & st-bce        & \textbf{83.8} & 69.2         & 78.8
                                      & 92.1 & 85.4         & 90.4
                                      &   –  &    –         &   –   \\
    \midrule

    \multicolumn{2}{l|}{\cite{foscarin2024beatthis}}  & 80.0 & 63.6 & 75.1
                                                     & 93.7 & 87.1 & 89.2
                                                     &   –  &   –  &   –   \\
    \multicolumn{2}{l|}{\cite{Bck2020DeconstructAR}} & 72.2 & 69.6 & 87.2
                                                     &  0.0 &  0.0 &  0.0
                                                     &   –  &   –  &   –   \\
    \multicolumn{2}{l|}{\cite{Hung2022}}                    & 74.8 & 73.8 & 87.0
                                                     & \textbf{94.5} & \textbf{93.9} & \textbf{95.9}
                                                     &   –  &   –  &   –   \\
    \multicolumn{2}{l|}{\cite{ZhaoXW22}}             & 74.8 & 71.2 & 84.1
                                                     &  0.0 &  0.0 &  0.0
                                                     &   –  &   –  &   –   \\

    \bottomrule
  \end{tabular}
  \end{sc}
}
\vskip -0.1in
\label{tab:cv_downbeat}
\end{table}

\end{document}